\documentclass[reprint,prl,twocolumn,superscriptaddress,showpacs]{revtex4-1}
\usepackage{graphicx} %insertion of eps figures
\usepackage{natbib} %implementation of \cite command; typical styles are apsrev is used for all the aps journals except rmp
\usepackage[usenames,dvipsnames]{color} % use colored text in latex
\usepackage{amsmath} % Defines extra environments for multiline displayed equations, as well as a number of other enhancements for math
\usepackage{amssymb} % {bm}% bold math
\usepackage{soul} %for underlining compatible with text wrapping; also other things
\usepackage{ifthen} %allows defining if then function
\newcommand{\dg}{^\dagger}

\newcommand{\rt}[1]{\sqrt{#1}}

\def\be#1\ee{\begin{equation}#1\end{equation}}
\def\ba#1\ea{\begin{align}#1\end{align}}
\def\bg#1\eg{\begin{gather}#1\end{gather}}

\newcommand{\abs}[1]{\lvert#1\rvert}

\def\shownote{0} %make this flag zero if you want text under \note to be shown
\newcommand{\note}[1]{\ifthenelse{\shownote=1}{\textcolor{Red}{[[#1]]}}{}}
\newcommand{\nt}[1]{\ifthenelse{\shownote=1}{\textcolor{Red}{[[#1]]}}{}}

\def\shownoteforauthors{0} %make this flag zero if you want text under \note to be shown
\newcommand{\nta}[1]{\ifthenelse{\shownoteforauthors=1}{\textcolor{Blue}{[[#1]]}}{}}
\def\showaddmat{0} %make this flag zero if you want text under \personote to be shown
\newcommand{\addmat}[1]{\ifthenelse{\showaddmat=1}{\textcolor{Gray}{[[#1]]}}{}}

\begin{document}

\title{Flux qubits in a planar circuit quantum electrodynamics architecture: quantum control and decoherence}

\author{J.-L. Orgiazzi \footnotemark[1] \footnotetext[1]{Corresponding author: orgiazzi@uwaterloo.ca}}
\email{orgiazzi@uwaterloo.ca}
\affiliation{Institute for Quantum Computing, Department of Electrical and Computer Engineering, and Waterloo Institute for Nanotechnology, University of Waterloo, Waterloo, ON, Canada N2L 3G1}

\author{C. Deng}
\affiliation{Institute for Quantum Computing, Department of Physics and Astronomy, and Waterloo Institute for Nanotechnology,
University of Waterloo, Waterloo, ON, Canada N2L 3G1}

\author{D. Layden}
\altaffiliation[Current address: ]{Department of Applied Mathematics, University of Waterloo, Waterloo, ON, Canada N2L 3G1}
\affiliation{Institute for Quantum Computing, Department of Physics and Astronomy, and Waterloo Institute for Nanotechnology,
 University of Waterloo, Waterloo, ON, Canada N2L 3G1}

\author{R. Marchildon}
\altaffiliation[Current address: ]{The Edward S. Rogers Department of Electrical and Computer Engineering, University of Toronto, 10 King\textquoteright{}s College Road, Toronto, ON, Canada M5S 3G4}
\affiliation{Institute for Quantum Computing, Department of Physics and Astronomy, and Waterloo Institute for Nanotechnology,
University of Waterloo, Waterloo, ON, Canada N2L 3G1}

\author{F. Kitapli}
\affiliation{Institute for Quantum Computing, Department of Electrical and Computer Engineering, and Waterloo Institute for Nanotechnology, University of Waterloo, Waterloo, ON, Canada N2L 3G1}

\author{F. Shen}
\affiliation{Institute for Quantum Computing, Department of Physics and Astronomy, and Waterloo Institute for Nanotechnology,
University of Waterloo, Waterloo, ON, Canada N2L 3G1}

\author{M. Bal}
\altaffiliation[Current address: ]{Materials Institute, T\"UBITAK Marmara Research Center, 41470 Gebze, Kocaeli, Turkey}
\affiliation{Institute for Quantum Computing, Department of Physics and Astronomy, and Waterloo Institute for Nanotechnology,
University of Waterloo, Waterloo, ON, Canada N2L 3G1}

\author{F. R. Ong}
\altaffiliation[Current address: ]{Institut f\"{u}r Experimentalphysik, Universit\"{a}t Innsbruck, Technikerstrasse 25/4, 6020 Innsbruck, Austria}
\affiliation{Institute for Quantum Computing, Department of Physics and Astronomy, and Waterloo Institute for Nanotechnology,
University of Waterloo, Waterloo, ON, Canada N2L 3G1}

\author{A. Lupascu \footnotemark[1] \footnotetext[1]{Corresponding author: alupascu@uwaterloo.ca}}
\affiliation{Institute for Quantum Computing, Department of Physics and Astronomy, and Waterloo Institute for Nanotechnology,
University of Waterloo, Waterloo, ON, Canada N2L 3G1}
\email{alupascu@uwaterloo.ca}

\date{ \today}

\begin{abstract}
%ABSTRACT
%\emph{Abstract version 1}
We report experiments on superconducting flux qubits in a circuit quantum electrodynamics (cQED) setup. Two qubits, independently biased and controlled, are coupled to a coplanar waveguide resonator. Dispersive qubit state readout reaches a maximum contrast of $72\,\%$. We find intrinsic energy relaxation times at the symmetry point of $7\,\mu\text{s}$ and $20\,\mu\text{s}$ and levels of flux noise of $2.6\,\mu \Phi_0/\rt{\text{Hz}}$ and $2.7\,\mu \Phi_0/\rt{\text{Hz}}$ at 1 Hz for the two qubits. We discuss the origin of decoherence in the measured devices. These results demonstrate the potential of cQED as a platform for fundamental investigations of decoherence and quantum dynamics of flux qubits.

\end{abstract}
%below there are some pacs that we would typically use, pls search everytime to find most suitable
\pacs{85.25.Cp%Josephson devices
, 42.50.Dv %Quantum state engineering and measurements
%, 03.65.Ta%Measurement theory (quantum mechanics)
%, 85.25.Dq%SQUIDs
, 03.67.Lx%Quantum computation
%07.57.Kp%microwave detectors~\cite{taylor_2008_NVDiamondMagn}
%, 74.50.+r %Tunneling phenomena; Josephson effects in superconductivity
, 74.78.Na 	%Mesoscopic and nanoscale systems
%85.25.Oj%superconducting photodetectors
}
\maketitle

Superconducting qubits are one of the main candidates for the implementation of quantum information processing~\cite{Devoret_2013_SupCircuits} and a rich testbed for research in quantum optics, quantum measurement, and decoherence~\cite{you_2011_revSupQbAMO}. Among various types of superconducting qubits, flux-type superconducting qubits have unique features. Strong and tunable coupling to microwave fields enables fundamental investigations in quantum optics~\cite{niemczyk_2010_cqedstrong,Forn-Diaz2010,Peropadre_2013_USC} and relativistic quantum mechanics~\cite{sabin_2012_PastFutureCorrelations}. The large magnetic dipole moment is a key ingredient in flux noise measurements \cite{Bylander2011}, sensitive magnetic field measurements \cite{Bal2012}, microwave-optical interfaces~\cite{Zhu_2011_FluxQubitNV}, and hybrid systems formed with nanomechanical resonators~\cite{Harrabi_2013_FluxQubit}. Finally, flux qubits have a large degree of anharmonicity which is an advantage for fast quantum control \cite{Groot2012}. Progress on these diverse research avenues has been hampered by relatively low and irreproducible coherence times compared to other types of superconducting qubits.

In the last decade, circuit quantum electrodynamics (cQED)~\cite{Blais2007,wallraff_2004_1} has become increasingly popular. In cQED, resonators provide a controlled electromagnetic environment protecting qubits from energy relaxation. In addition, resonators are used for qubit state measurement~\cite{wallraff_2005_1} and as quantum buses for qubit-qubit coupling~\cite{Majer2007}. In this letter, we present an implementation of cQED with flux qubits strongly coupled to a superconducting coplanar waveguide resonator. The qubits and the resonator are made of aluminum. Local biasing and control lines provide a mean to implement fast single qubit gates as well as controlled two-qubit interactions. We measure energy relaxation times around $10\,\mu\text{s}$, an improvement over previous experiments with flux qubits coupled to coplanar waveguide resonators~\cite{Inomata2012,Jerger2012}, and comparable with the longest measured to date on flux qubits~\cite{Bylander2011,bertet_2005_1}. We characterize in detail the decoherence of the flux qubits coupled to the resonator. Based on decoherence measurements, we extract levels of flux noise of $2.6\,\mu \Phi_0/\rt{\text{Hz}}$ and $2.7\,\mu \Phi_0/\rt{\text{Hz}}$ at 1 Hz for the two qubits. We also present a spectroscopic measurement of a resonator-mediated qubit-qubit coupling, which is relevant for implementation of two-qubit gates. These results demonstrate the versatility of cQED with flux qubits, and its potential for further understanding and improvements of decoherence of these qubits.

%INSERTING FIG 1
\begin{figure*}[!]
\includegraphics[width=17.6cm]{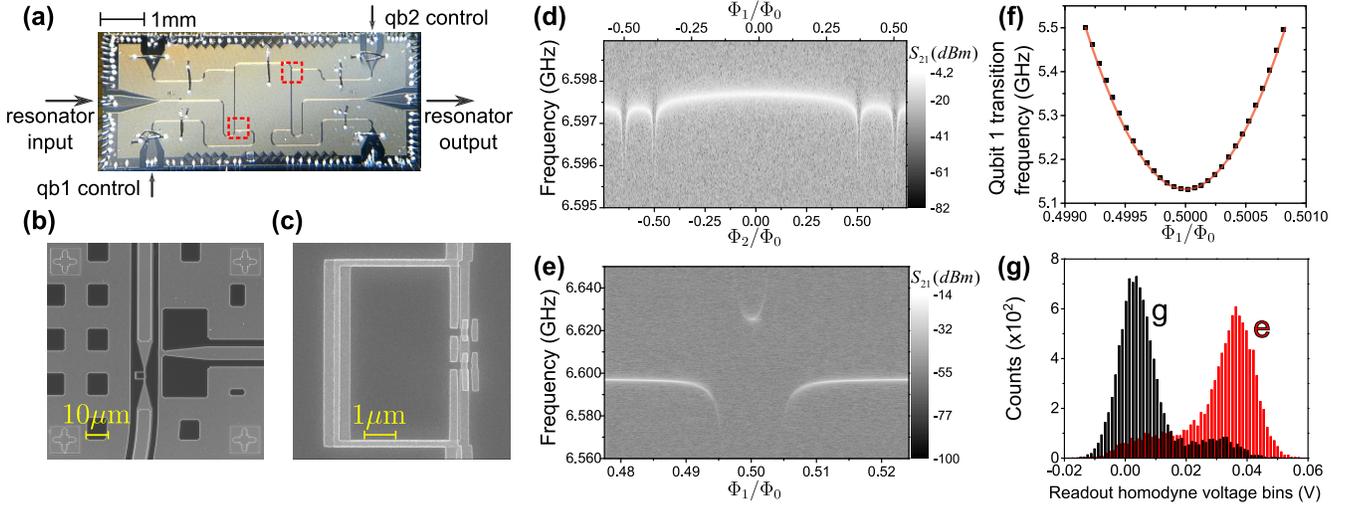}
\caption{\label{fig:fig1}
\textbf{(a)} Optical image of the device where the overlaid dashed rectangles indicate the position of qubit 1 (bottom left) and qubit 2 (top right). \textbf{(b)} Scanning electron microscope (SEM) image showing a qubit embedded in the CPW resonator and its local CPW control line. \textbf{(c)} SEM image of a qubit device nominally identical to that used in this work. \textbf{(d)} Resonator transmission (greyscale) versus frequency (vertical axis) and flux bias of qubit 1/2 (top/bottom axis). \textbf{(e)} Resonator transmission (greyscale) versus frequency and flux bias of qubit 1. \textbf{(f)} Qubit 1 spectroscopy measurements. The transition frequency is plotted versus the applied magnetic flux bias. The continuous line is a fit of the persistent current qubit model, yielding $I_{\rm p1}=383\,\text{nA}$ and $\Delta_{1}=5.1317\,\text{GHz}$. \textbf{(g)} Readout histograms for qubit 1 resulting in a qubit state readout contrast of $72\,\%$.}
\end{figure*}

The device used in our work is shown in Fig.~\ref{fig:fig1}(a). It contains a coplanar waveguide (CPW) resonator, with two ports used for microwave transmission measurements. Two qubits are coupled to the CPW resonator, via the mutual inductance of a shared line (Fig.~\ref{fig:fig1}(b)). The qubits are persistent current type flux qubits \cite{Mooij1999}, consisting of a superconducting loop interrupted by four Josephson junctions (Fig.~\ref{fig:fig1}(c)). A CPW line terminated by a low inductance shunt is coupled to each qubit and used to send microwave pulses for coherent qubit control (see Fig.~\ref{fig:fig1}(b)). The device is fabricated on an intrinsic silicon substrate, in a two step process. In the first step, optical lithography, evaporation of a 190~nm thick aluminum layer and liftoff are used to define the resonator and the control lines. In the second step, a bilayer resist is patterned using electron-beam lithography. Subsequently, shadow evaporation of two aluminum layers, 40 and 65 nm thick respectively, followed by liftoff define the qubit junctions. An argon milling step, done before the shadow evaporation, is critical to ensure a high quality contact between the two aluminum layers and for the reproducibility of the Josephson junctions.

Experiments are performed in a dilution refrigerator, using a custom-designed probe for microwave transmission measurements \cite{Ong2012}. The chip is enclosed in a copper box, which is placed inside a three-layer high permeability metal shield. An active magnetic field compensation system placed outside the cryostat is used to further reduce low-frequency magnetic field noise. A set of superconducting coils, attached to the device copper box, is used to provide independent magnetic flux biases to the two qubits. Qubit state control is done using shaped microwave pulses. Qubit state measurement is done in the dispersive regime~\cite{wallraff_2005_1}, by measuring the transmission of microwave pulses through the resonator. The transmission lines for qubit control and readout are filtered using attenuators and filters placed at different temperature stages. A detailed description of the experimental setup is provided in the Supplemental Material.

We first discuss the model describing the qubits and the resonator. When the magnetic flux $\Phi_i$ ($i=1,2$) applied to qubit $i$ is close to $\overline{\Phi}_{n}=(n+1/2)\Phi_0$, with $n$ an integer and $\Phi_0=h/2e$ the flux quantum, the qubit is described by the Hamiltonian $H_{\text{qb,}i}=-\varepsilon_i/2 \sigma_{z,i}-h\Delta_i/2 \sigma_{x,i}$. Here $\varepsilon_i=2I_{\rm pi}(\Phi_i-\overline{\Phi}_{n_i})$, where $n_i$ ($i=1,2$) are integers, and $I_{\rm pi}$ and $\Delta_i$, called the persistent current and gap~\cite{orlando_1999_1}, are determined by the qubit design parameters. The operators $\sigma_{z/x,i}$, $i=1,2$, are the Pauli Z and X operators for qubit $i$. The resonator Hamiltonian is $H_\text{res}=\sum_{j\geq 1} \hbar \omega_{r,j} a_j\dg a_j$, with $\omega_{r,j}$ the resonance frequencies and $a_j\dg$ ($a_j$) the creation (annihilation) operator for mode $j$. The interaction between qubit $i$ and mode $j$ of the resonator is given by $H_{\text{int},ij}=h g_{i,j}\sigma_{x,i}(a_j\dg+a_j)$, with $g_{i,j}$ coupling factors. Both qubits are very strongly coupled to the resonator, making it important to account for multiple resonator modes and keep counter-rotating terms.

 We first present experiments on the spectroscopic characterization of the coupled qubit-resonator system. A continuous wave transmission measurement of the resonator, taken versus the applied magnetic field, is shown in Fig.~\ref{fig:fig1}(d). We observe the resonance corresponding to the first mode of the resonator at $\omega_{r,1}=2\pi\times 6.597\,\text{GHz}$. A significant change in the response occurs when the flux through each qubit is close to $-\Phi_0/2$ and $\Phi_0/2$. A narrower range scan of the transmission for qubit 1, done with a power corresponding to an average of 0.6 photons in the resonator, is shown in Fig.~\ref{fig:fig1}(e). An anticrossing is observed where the qubit and cavity are resonant. Next, qubit spectroscopy is performed by applying microwave pulses to each qubit local CPW control line. In Fig.~\ref{fig:fig1}(f) we show the spectroscopically measured transition frequency for qubit 1 versus magnetic flux. These data and similar data obtained for qubit 2 (not shown), are used to extract the qubit parameters $I_{\rm p1}=383\,\text{nA}$, $\Delta_{1}=5.1317\,\text{GHz}$, $I_{\rm p2}=352\,\text{nA}$, and $\Delta_{2}=3.6634\,\text{GHz}$. For each qubit, we use measurements of photon number splitting~\cite{Schuster2007a} for photons populating the first mode, together with a model which takes into account the first ten modes of the CPW resonator, to extract the coupling to the first resonator mode $g_{1,1}=155.6\,\text{MHz}$ and $g_{2,1}=295.4\,\text{MHz}$.

 Qubit state readout is performed using homodyne detection~\cite{wallraff_2005_1}. To optimize the readout contrast, the cavity is driven strongly, in the nonlinear regime. A histogram of the homodyne voltage for qubit 1, averaged over a readout pulse duration of $4\,\mu s$, is shown in Fig.~\ref{fig:fig1}(g). The readout contrast for this qubit is $72\,\%$. Similar results (not shown) are obtained for qubit 2, where the maximum readout contrast is $60\,\%$. For both qubits, readout contrast is limited primarily by the initialization procedure which is based on thermalization.

\begin{figure}[!]
\includegraphics[width=7.4cm]{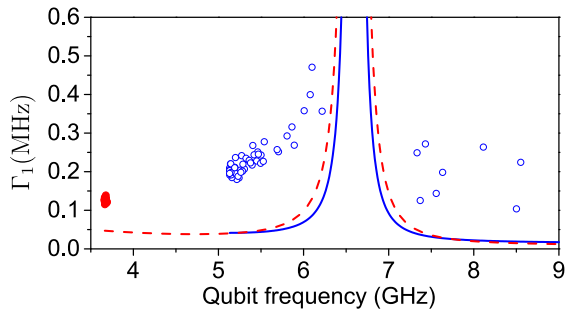}
\caption{\label{fig:fig2} Energy relaxation rate versus transition frequency for qubit 1 (open dots) and 2 (closed dots). The combined energy relaxation rate due to the resonator and the CPW control line is shown by the continuous (dashed) line for qubit 1\,(2).}
\end{figure}

\begin{figure}[!]
\includegraphics[width=8.6cm]{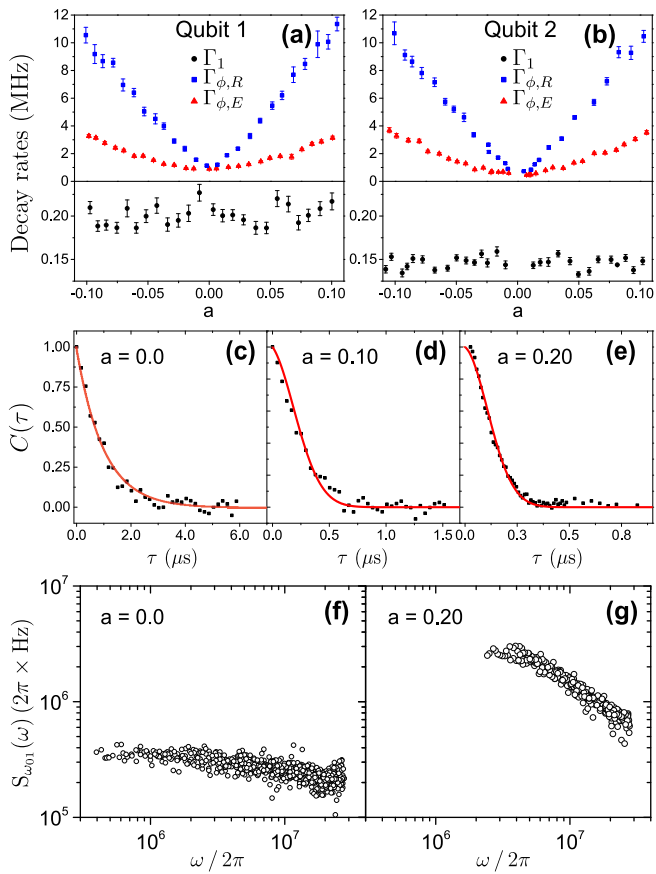}
\caption{\label{fig:fig3} \textbf{(a, b)} Energy relaxation (black dots) and dephasing rates performed using Ramsey (blue squares) and spin-echo pulse (red triangles) sequences for qubits 1 (a) and 2 (b). \textbf{(c, d, e)} Spin-echo decay for qubit 1 for different coupling angles to flux noise, where the solid lines represent a fit to a coherence function defined in the text. \textbf{(f, g)} Qubit 1 frequency noise PSD calculated from measurements based on dynamical decoupling at the symmetry point (f) and at a coupling angle $a=0.2$ (g).}
\end{figure}

 We next present energy relaxation measurements. The energy relaxation times are $T_{1}=5.3\,\mu\text{s}$ and $9.6\,\mu\text{s}$ for qubits 1 and 2 respectively at their symmetry points. A measurement of energy relaxation rates versus magnetic flux around the symmetry point is shown in Figs.~\ref{fig:fig3}(a) and (b). Over the explored frequency range, of $35.6\,\text{MHz}$ and $24.9\,\text{MHz}$ for qubits 1 and 2 respectively, we observe only minor variations of the energy relaxation rate, of less than 15\% between extreme values. In Fig.~\ref{fig:fig2} we show the energy relaxation rate $\Gamma_1$ for qubits 1 and 2 over a broad range, together with a plot of the calculated rate induced by the electromagnetic environment. The latter takes into account relaxation through Purcell effect~\cite{Houck2008} due to the first 10 modes of the resonator and relaxation due to the control line. If the relaxation due to these sources is excluded, we calculate intrinsic relaxation times of $7\,(20)\,\mu\text{s}$  for qubit 1\,(2). In a related work we considered the role of quasiparticles in persistent current qubits~\cite{bal_2014_tlf}. The measured intrinsic energy relaxation times can be attributed to a non-equilibrium quasiparticle density of $0.12\,\mu m^{-3}$ and $0.04\,\mu m^{-3}$ respectively, in line with other measurements on similar devices (see~\cite{bal_2014_tlf} and references therein). While quasiparticles are the main candidate for energy relaxation, we do not exclude other potential sources, in particular loss due to amorphous interfaces and surfaces~\cite{Barends_2013_Xmon}. We note that lower energy relaxation times, in the $0.5-1\,\mu\text{s}$ range, were obtained in previous experiments with aluminum flux qubits coupled to superconducting resonators made of niobium~\cite{Inomata2012,Jerger2012}. Possible reasons for the longer relaxation times in our experiment include a reduction of quasiparticle induced relaxation, due to using an all aluminum circuit, and a reduction of surface/interface loss arising due to the different processing prior to deposition of the qubit layer.

%INSERTING FIG 2

We now turn to a discussion of dephasing. In Figs.~\ref{fig:fig3}(a) and (b) we present detailed measurements of dephasing performed using Ramsey and spin-echo pulse sequences~\cite{Ithier2005}. Away from the symmetry point, the increased sensitivity to magnetic flux renders flux noise the dominant contribution to decoherence. We fit the coherence decay over the time $\tau$ using the expression $e^{-\Gamma_1\tau/2}e^{-(\Gamma_{\phi}\tau)^2}$~\cite{Ithier2005}, with $\Gamma_1$ the energy relaxation rate and $\Gamma_{\phi}$ the pure dephasing rate. The latter depends on the type of experiment; for Ramsey (spin-echo) measurements, we denote this rate by $\Gamma_{\phi,R}$ ($\Gamma_{\phi,E}$). Gaussian decay is predicted when decoherence is dominated by noise with a power spectral density (PSD) proportional to $\abs{\omega}^{-1}$\cite{Ithier2005,Kakuyanagi2007,Yoshihara2006}, with $\omega$ the angular frequency. Assuming flux noise with a PSD given by $A/\abs{\omega}$, the slope of $\Gamma_{\phi,E} (a)$, with  $a=\varepsilon/\hbar \omega_{01}$, where $\omega_{01}$ is the angular transition frequency can be used to determine $A$~\cite{Yoshihara2006}. We find $\sqrt{A}=2.6\,(2.7)\,\mu\Phi_0$ for qubit 1\,(2). These levels of flux noise are slightly larger than for the smaller aluminum flux qubits in Refs.~\onlinecite{Yoshihara2006,Bylander2011}, in qualitative agreement with size scaling~\cite{Anton_2013_FluxNoiseSquidsTdep}.

%INSERTING FIG 3

Next we discuss the dephasing of the qubits close to the symmetry point. At the symmetry point, decay is exponential, with rates $\overline{\Gamma}_{\phi,R}^{-1}= 0.77\,(0.90)\,\mu\text{s}$ for Ramsey and $\overline{\Gamma}_{\phi,E}^{-1}=1.03\,(1.85)\,\mu\text{s}$ for spin-echo measurements for qubit 1\,(2). We observe that both Ramsey and spin-echo curves change shape from exponential at the symmetry point to Gaussian away from this point (see Figs.~\ref{fig:fig3}(c)-(e) for spin-echo measurements of qubit 1). This suggests that dephasing can be explained by the combination of an exponential decay process and a Gaussian decay process. The latter is due to magnetic flux noise and has a rate $\Gamma_{\phi,R/E} = \gamma_{R/E} a$, for Ramsey/spin-echo, with $\gamma_{R/E}$ dependent on qubit parameters and flux noise amplitude~\cite{Yoshihara2006}. Indeed, we find that all the coherence decay curves for each qubit can be fit by the expression $C(\tau)=e^{-\Gamma_1\tau/2}e^{-\overline{\Gamma}_{\phi R/E}\tau} e^{-(\gamma_{R/E} a \tau)^2}$, for Ramsey/spin-echo, with $\gamma_{R/E}$ as a single fit parameter (see Figs.~\ref{fig:fig3}(c)-(e)).We also performed noise measurements based on dynamical decoupling \cite{Bylander2011}, shown in Figs.~\ref{fig:fig3}(f) and (g) for qubit 1 biased at $a=0$ (the symmetry point) and at $a=0.2$ respectively. This additional experiment confirms that a nearly frequency independent white noise source dominates dephasing at the symmetry point.

We discuss next the origin of the decoherence at the symmetry point. We first consider quadratically coupled flux noise. As discussed in Ref.~\onlinecite{makhlin_2004_1}, the decay is expected to be significantly non-exponential at short time, with a time scale estimated to be $33\,(42)\,\mu s$ for qubit 1 (2). We have also performed numerical simulations that confirm this source is negligible. A second potential source is photon noise induced dephasing \cite{Schuster2005,bertet_2005_1}. We performed spectroscopy experiments (not shown) and numerical simulations which allow us find an upper bound for the thermal photon number $n_{\text{th}}<0.02$. In the strong dispersive regime~\cite{rigetti_2012_0p1msqubit,sears_2012_PhotTrans} the dephasing rate, given by $\kappa n_\text{th} $, with $\kappa = 882\,\text{kHz}$ the resonator decay rate for the first mode, is smaller than $18\,\text{kHz}$, thus a  negligible contribution to dephasing.

We next consider dephasing due to charge noise, arising either from offset charges or quasiparticles on the qubit islands, as a potential source of dephasing at the symmetry point. The modulation of the persistent qubit transition frequency by charges, denoted by $\delta\Delta_\text{c}$, decreases exponentially with the ratio of the Josephson ($E_J$) to charging ($E_c$) energy~\cite{orlando_1999_1}. We numerically calculate $E_J$ and $E_c$ assuming proportional and inversely proportional respectively scaling with Josephson junction areas, as measured for a nominally identical device, and using the experimentally measured persistent current and gap. The thus estimated values of $E_J$ and $E_c$ yield $\delta\Delta_\text{c}=83~(52)\,\text{kHz}$ for qubit 1\,(2). However we note that this value is strongly dependent on the junction areas; assuming a size of the smallest junction different by only 10\%, a difference that could arise due to lithography or edge effects, we find $\delta\Delta_\text{c}=4~(3)\,\text{MHz}$ for qubit 1\,(2). With these larger values, dephasing could arise through a combination of slow offset charge fluctuations and random telegraph noise due to quasiparticles tunneling with a rate larger than the charge modulation.

%INSERTING FIG 4
\begin{figure}[!]
\includegraphics[width=7.5cm]{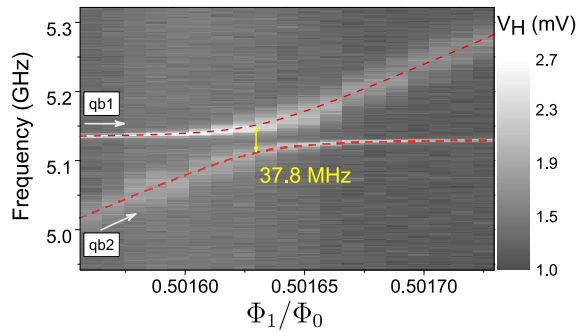}
\caption{\label{fig:fig4} Readout homodyne voltage ($V_{\rm H}$) versus frequency (vertical axis) and the flux applied to qubit 2. The overlaid dashed lines are a fit of the excited energies for the coupled system, providing a coupling strength $37.8\,\text{MHz}$.}
\end{figure}

We finally discuss the use of the proposed setup to implement qubit-qubit interactions. In Fig.~\ref{fig:fig4} we show spectroscopy measurements where qubit 1 is biased at the symmetry point, whereas the flux bias of qubit 2 is changed. An anticrossing arises due to an effective qubit-qubit interaction, mediated by virtual excitations of the resonator~\cite{Blais2007}. The large qubit-qubit interaction of $37.8\,\text{MHz}$ is made possible by the strong coupling of the flux qubits to the resonator. A two-qubit gate can be implemented using the method proposed in~\cite{Groot2010}, which only requires control of each qubit, a feature already included in our setup. The large anharmonicity of the flux qubits enables fast two-qubit gates, with a time limited by the inverse of the interaction strength~\cite{Groot2012}. We will present these results in a follow-up paper. The ability to perform high-fidelity gates between distant qubits will be important for experiments using flux qubits in hybrid architectures~\cite{Zhu_2011_FluxQubitNV,Harrabi_2013_FluxQubit}.

We presented experiments on flux qubits coupled to a superconducting on-chip resonator. The measured qubits have long energy relaxation times and low levels of flux noise. Readout contrast is high, exceeding 70\%. We also demonstrated the strong, resonator mediated, interaction between the two qubits. Further improvements of coherence will have to address the role of quasiparticles and loss due to surfaces and interfaces in energy relaxation and the origin of pure dephasing at the flux-insensitive point. The experiments presented here demonstrate the potential that this platform has for systematic studies of coherence and dynamics of flux qubits.

%ACKNOWLEDGEMENTS
We acknowledge useful discussions with Patrice Bertet and Pol Forn-D\'{\i}az. We acknowledge support from NSERC through Discovery and RTI grants, Canada Foundation for Innovation, Ontario Ministry of Research and Innovation, Industry Canada, and CMC Microsystems. During this work, AL was supported by a Sloan Research Fellowship, CD was supported by an Ontario Graduate Scholarship, and DL and RM were supported by NSERC USRA scholarships.

\emph{Note.} A recent paper~\cite{stern_2014_3dPCQ} reports on related experiments on measurements of decoherence of persistent current qubits coupled to a three-dimensional resonator.

\clearpage
\pagebreak
\widetext
\begin{center}
\textbf{\large Supplementary information: Flux qubits in a planar circuit quantum electrodynamics architecture: quantum control and decoherence}
\end{center}
%%%%%%%%%% Merge with supplemental materials %%%%%%%%%%
%%%%%%%%%% Prefix a "S" to all equations, figures, tables and reset the counter %%%%%%%%%%
\setcounter{equation}{0}
\setcounter{figure}{0}
\setcounter{table}{0}
\setcounter{page}{1}
\makeatletter
\renewcommand{\theequation}{S\arabic{equation}}
\renewcommand{\thefigure}{S\arabic{figure}}
\renewcommand{\bibnumfmt}[1]{[S#1]}
\renewcommand{\citenumfont}[1]{S#1}
%%%%%%%%%% Prefix a "S" to all equations, figures, tables and reset the counter %%%%%%%%%%

\section{Experimental setup}

\begin{figure}[htp]
\includegraphics[width=17.8cm]{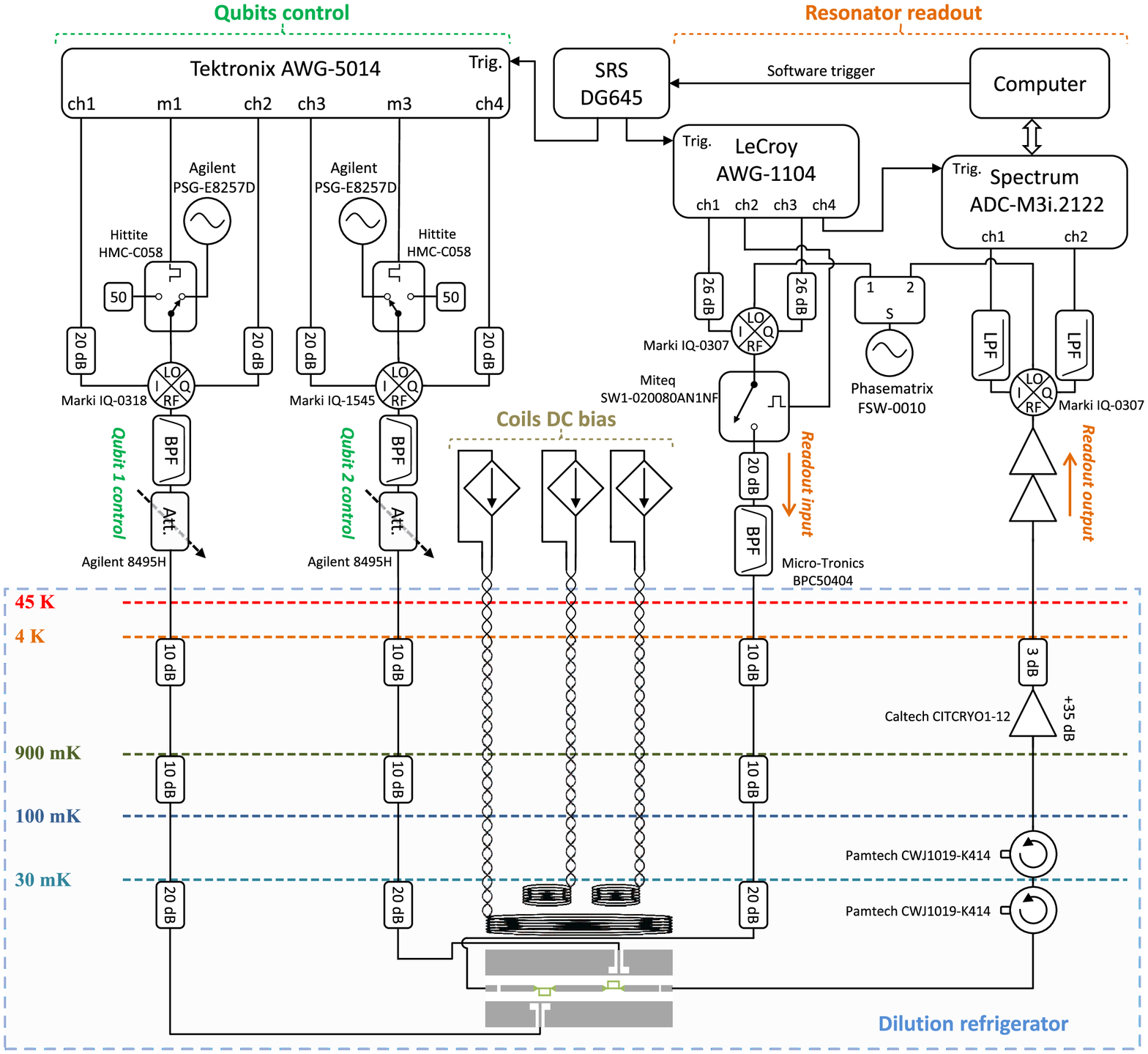}
\caption{\label{fig:S_fig1}
\textbf{Detailed schematic of the experimental setup.} The description is given in the text.}
\end{figure}

Experiments are performed in a Leiden Cryogenics dilution refrigerator, model CF-650. We use a custom design top loading probe for microwave transmission measurements~\cite{Ong2012}. Supplementary Figure~\ref{fig:S_fig1} shows a detailed schematic of the experimental setup. The chip containing two flux qubits and a coplanar waveguide resonator is enclosed in a copper box with low mode volume, which is placed inside a three-layer high permeability metal shield. An active magnetic field compensation system placed outside the cryostat is used to further reduce slow drifts of the ambient magnetic field. A set of superconducting coils is attached to the device copper box, and used to provide independent magnetic flux biases to the two qubits. The coils are supplied by custom designed ultra-stable voltage to current converters, each controlled by a Yokogawa 7651 voltage source. The temperature measured at the mixing chamber level of the custom designed top loading probe, to which the device copper box is attached, was 40 mK.

Each qubit is individually controlled using shaped microwave pulses that are generated using a microwave synthesizer Agilent PSG E8257D, an arbitrary waveform generator Tektronix AWG5014, and a Marki IQ-0318 (IQ-1545) mixer for qubit 1 (2). To further reduce microwave leakage to the qubit during coherent state evolution or qubit ground state initialization, each microwave synthesizer is isolated from the corresponding mixer by a Hittite HMC-C058 switch which provides an isolation $\geq65$ dB up to 6 GHz. Agilent 8495H Programmable step attenuators are used to adjust the pulse amplitude over a wide range. A band pass filter (BPF) in each control line is used to suppress low-frequency noise and spurious harmonics, which are detrimental to qubit coherence, and to prevent cavity excitation.

%INSERTING FIG 1

Qubit state measurement is done using dispersive readout \cite{wallraff_2005_1}, by measuring the transmission of microwave pulses through the resonator, using a custom built setup. The readout pulses are generated in a similar fashion to the qubit control pulses, using a Phase Matrix QuickSyn FSW-0010 microwave synthesizer, a LeCroy AWG-1104 arbitrary waveform generator, a Marki IQ-0307 mixer, and a Miteq switch model SW1-020080AN1NF. The readout input line is filtered by a 6 -10 GHz band pass filter from Micro-Tronics, model BPC50404. The readout output signal is amplified using cryogenic and room temperature amplifiers. After demodulation, done with another Marki IQ-0307 mixer, the quadratures of the readout output pulse are sampled using a Spectrum M3i.2122 digitizer. A Stanford Research Systems digital delay generator DG645 controls the synchronization of the arbitrary waveform generators and digitizers.

All the transmission lines inside the cryostat are coaxial lines made of either beryllium copper or stainless steel. Microwave attenuators, from XMA, are placed at different temperature stages. In addition, the output readout line contains two Pamtech CWJ1019-K414 isolators thermally anchored at the mixing chamber plate, and a cryogenic amplifier Caltech CITCRYO1-12 at the 4 K plate.

\section{Dynamical decoupling pulse sequence}
\begin{figure}[htp]
\includegraphics[width=12cm]{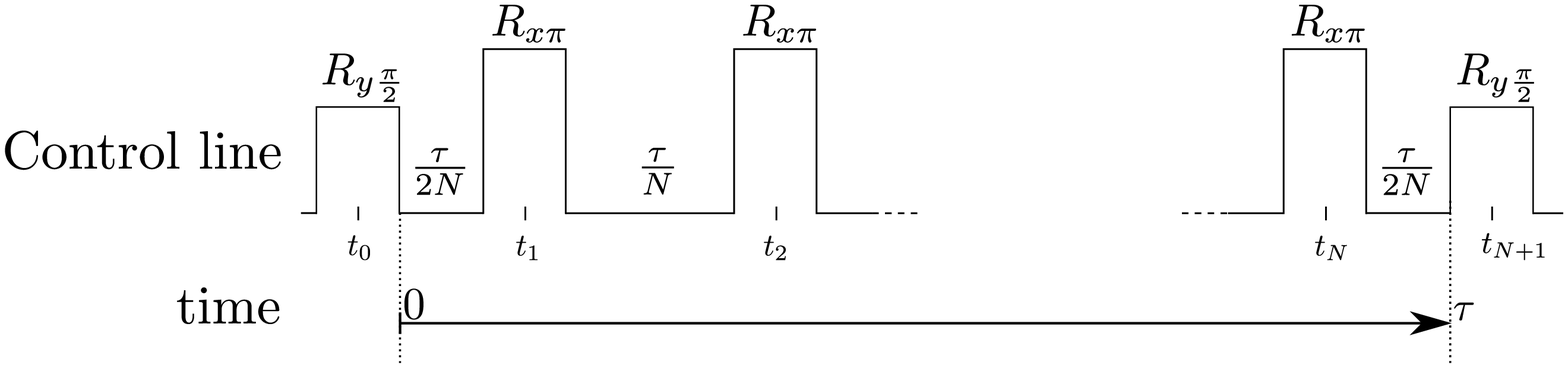}
\caption{\label{fig:S_fig2}
\textbf{Carr-Purcel-Meiboom-Gill (CPMG) pulse sequence.} CPMG pulse sequence used to measure the qubit transition frequency power spectral density. The sequence contains N $\pi$ rotations around the $x$ axis, denoted by $R_{x\pi}$, included between $\pi/2$ rotations around the $y$ axis, denoted by $R_{y\pi/2}$.
}
\end{figure}

In order to evaluate the qubit transition frequency noise power spectral density (PSD), we use a Carr-Purcel-Meiboom-Gill (CPMG) pulse protocol and a post processing approach to compute the PSDs as presented in Refs. \onlinecite{Biercuk2009,Biercuk2011a,Bylander2011}. Well calibrated pulses are used. Supplementary Figure~\ref{fig:S_fig2} shows the pulse sequence, containing N $\pi$ pulses.

%INSERTING FIG 2

\begin{figure}[htp]
\includegraphics[width=14.0cm]{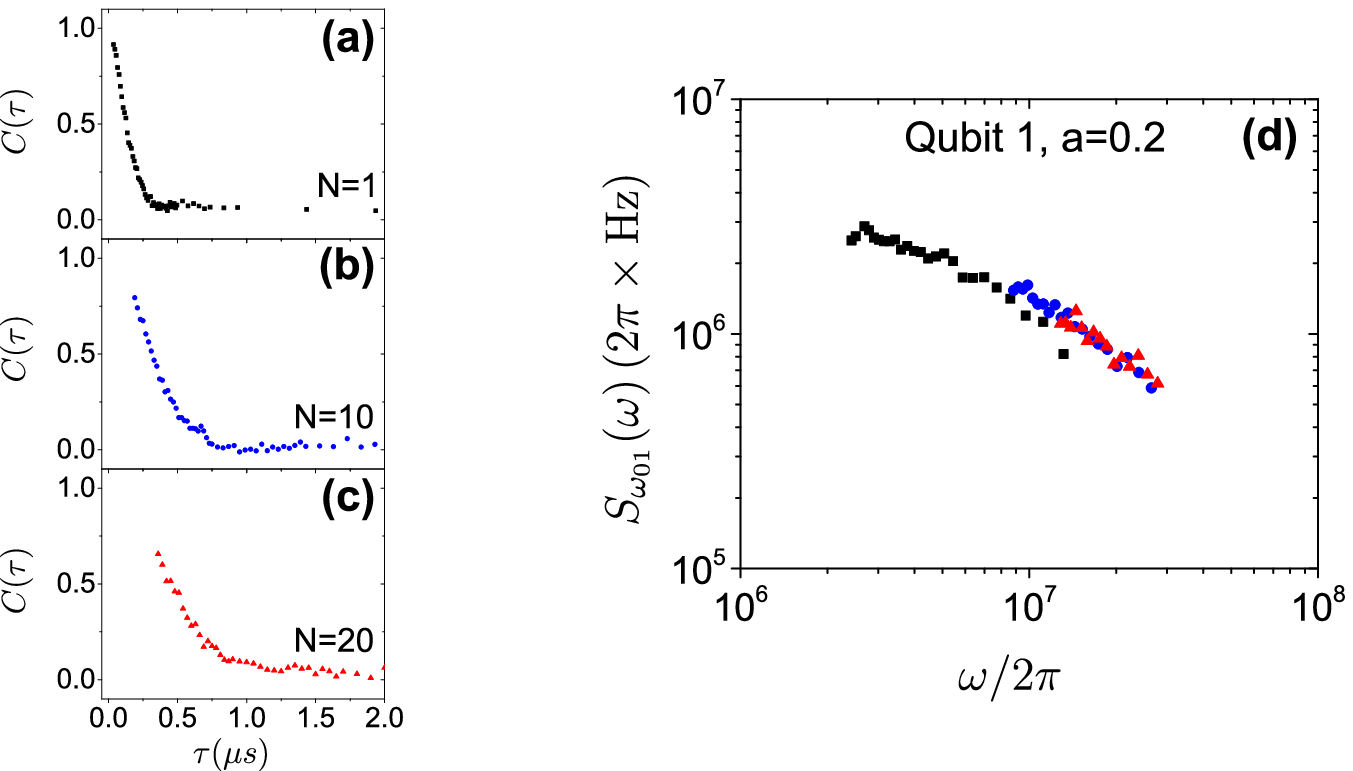}
\caption{\label{fig:S_fig3}
\textbf{Qubit 1 coherence decay with a CPMG pulse sequence and calculated frequency noise PSD.} \textbf{(a-c)} Qubit 1 coherence function $C(\tau)$, with $\tau$ the total time between the $\frac{\pi}{2}$ pulses, shown for three different values of N - the number of $\pi$ pulses.\textbf{(d)} Qubit 1 frequency noise power spectrum density, when biased away from symmetry point (a=0.2). The black squares are calculated from the coherence function data points in (a), the blue dots from (b) and the red triangles from (c).
}
\end{figure}

The decay of coherence for qubit 1 versus the evolution time is shown in Figure~\ref{fig:S_fig3}(a)-(c), for CPMG pulses with N taking values 1, 10, and 20. Figure~\ref{fig:S_fig3}(d) shows the calculated qubit frequency noise PSD from those decay curves (matching symbol and color between the decay points used to compute the PSD in (a)-(c) and (d)), where qubit 1 was far from its optimal bias point, at $a = 0.2$, where the transition frequency noise is due primarily to magnetic flux fluctuations. We find a transition frequency noise PSD which is well approximated by a $1/\omega^{\alpha}$ dependence, with $\alpha =0.71$.

%INSERTING FIG 3

\end{document}